\documentclass[aps, prb, twocolumn, showpacs, superscriptaddress, preprintnumbers, bibnotes, amsmath, amssymb]{revtex4-2}
\usepackage[utf8]{inputenc}
\usepackage{graphicx}
\graphicspath{{./mainfigure/}}
\usepackage{dcolumn}
\usepackage{bm}
\usepackage{adjustbox}
\usepackage{amssymb}
\usepackage{xcolor}

\usepackage[colorlinks=true, linkcolor=blue, anchorcolor=blue, citecolor=blue, urlcolor=blue]
{hyperref}
\begin{document}

\title{Tunable Dual-Type Weyl Points in Dirac-Weyl Semimetal CaAgBi}

\author{Shenghao Huang}
\affiliation{Department of Physics, Shanghai Key Laboratory of High Temperature Superconductors, Materials Genome Institute, International Centre of Quantum and Molecular Structures, Shanghai University, Shanghai 200444, China}

\author{Heng Gao}%
 \email{gaoheng@shu.edu.cn}

 \affiliation{Department of Physics, Shanghai Key Laboratory of High Temperature Superconductors, Materials Genome Institute, International Centre of Quantum and Molecular Structures, Shanghai University, Shanghai 200444, China}

 \author{Hongfei Wang}%
\email{hfwang@shu.edu.cn}
\affiliation{Department of Physics, Shanghai Key Laboratory of High Temperature Superconductors, Materials Genome Institute, International Centre of Quantum and Molecular Structures, Shanghai University, Shanghai 200444, China}

\author{Wei Ren}
\affiliation{Department of Physics, Shanghai Key Laboratory of High Temperature Superconductors, Materials Genome Institute, International Centre of Quantum and Molecular Structures, Shanghai University, Shanghai 200444, China}

\date{\today}

\begin{abstract}
Dirac-Weyl semimetals host both Dirac and Weyl fermions and the exploration of material candidates with tunable topological properties is essential to realize topological spintronic devices. In this work, we propose CaAgBi as a Dirac-Weyl semimetal with tunable type-I and type-II Weyl points based on first-principle calculations. In addition to the three pairs of Dirac points located along the rotational axis as previously reported, our calculations reveal 18 additional pairs of dual-type Weyl points distributed across three distinct planes: type-I in the $k_z=0$ plane and type-II in the $k_z= \pm 0.0698\,\frac{2\pi}{c}$ planes. The topological features are further confirmed through chirality of the Weyl points as well as the existence of surface Fermi arcs. Moreover, we find that the position and  annihilation of Dirac and Weyl points are tunable by the alloy engineering and external strains. The alloy engineering is employed to modulate the positions of Weyl points, revealing different annihilation concentrations for type-I and type-II Weyl points, potentially offering novel experimental strategies for Weyl point manipulation. Under tensile biaxial strain, the Weyl points in the $k_{z}=0$ plane annihilate along the $\Gamma$--$\mathrm{M}$ path at $2.1\%$ strain, whereas the Weyl points in the $k_{z}\neq 0$ planes remain robust within $6\%$ strain. This work provides a versatile platform for manipulating Dirac/Weyl interactions, with spin-orbit coupling (SOC) driven alloy control and strain-selective engineering opening avenues for topological electronics.
\end{abstract}

\maketitle


\section{\label{sec:level1}INTRODUCTION}

In recent years, topological semimetals have emerged at the forefront of condensed matter physics and materials science, due to their unique electronic structures and novel quantum phenomena. The electronic structures of topological semimetals typically exhibit linear or quasi-linear band crossings near the Fermi level, such as Dirac points, Weyl points, or nodal lines. Topological semimetals are classified based on the dimensionality (points, lines, surfaces) and degeneracy (twofold, fourfold) of these band-crossing features, leading to distinct categories mainly including Dirac semimetals \cite{young2012dirac,liu2014discovery,crassee20183d,yin2022electronic,ram2024magnetotransport,fang2023measurement,feng2022prediction}, Weyl semimetals \cite{zhang2018coexistence,lv2015experimental,du2023effect,wang2022ndalsi,lu2024realization,ono2021surface,kanagaraj2022topological,sun2015topological,weng2015weyl}, nodal-line semimetals \cite{kim2015dirac,regmi2024electronic,yu2015topological,su20224,du2023structural,wang2017topological,xu2018trivial}. In particular, centrosymmetric Dirac semimetals host Dirac points which are protected by both time-reversal ($T$) and inversion ($P$) symmetries, characterized by fourfold degenerate points near the Fermi level with linear dispersion. In contrast, a Weyl semimetal can be realized by breaking either $T$ or $P$ symmetry. Its band structure exhibits pairs of Weyl points near the Fermi level, characterized by linear dispersion and definite chirality. These paired Weyl points can merge into a Dirac point when they meet at the same $\mathbf{k}$-point, provided that symmetry allows. Such unique electronic structures give rise to various intriguing physical phenomena. For instance, Weyl semimetals exhibit Fermi arcs observable via angle-resolved photoemission spectroscopy (ARPES) \cite{lv2015experimental,damascelli2004probing}, chiral anomaly-induced negative magnetoresistance \cite{son2012berry,son2013chiral}, and the anomalous Hall effect \cite{liu2018giant} attributed to large Berry curvature.

Recently, Dirac-Weyl semimetals have been proposed to exist in polar hexagonal LiGaGe-type materials, such as SrHgPb \cite{gao2018dirac}, representing a novel topological phase characterized by the coexistence of Dirac and Weyl points. This unique topological phase offers an opportunity for investigating interactions between Dirac and Weyl fermions. Subsequent studies have identified other Dirac-Weyl semimetals, including SrHgSn, CaHgSn, as well as materials featuring two Weyl point planes such as MoTe and WTe \cite{meng2019dirac}. In this work, we present a systematic study of CaAgBi, a material already synthesized experimentally and found to crystallize in the hexagonal space group $\mathit{P}6_3\mathit{mc}$ (No.186) \cite{sun2007synthesis}, thus laying the groundwork for experimental exploration of its physical properties. Previous theoretical predictions have identified CaAgBi as a topological semimetal hosting three Dirac points, comprising both type-I and type-II \cite{chen2017ternary}. Given that CaAgBi shares the same space group and exhibits a similar low-energy band structure to SrHgPb, we expect that CaAgBi may also represent a Dirac-Weyl semimetal phase. 

Motivated by the above considerations, we constructed a tight-binding model to elucidate the topological properties of CaAgBi through analyses of the band structure, chirality of the Weyl points, Fermi arc surface states, and low energy effective model Hamiltonian \cite{liu2010model}. By examining the two-dimensional band structure near the Weyl points, we classified the types of Weyl points in each plane. Alloy engineering and external strain have emerged as widely adopted strategies in both experimental and theoretical studies for tuning the electronic properties of materials \cite{fang2020ideal,huang2018alloy,huang2014nontrivial,zubair2025weyl}. Taking advantage of alloy engineering, we controlled the doping concentration to achieve systematic modulation of the position of Weyl points. Biaxial strain and hydrostatic pressure were also applied to systematically tune the topological states and evaluate their robustness.

This paper is organized as follows. Section II is the computational method details for the first principles.  Section III comprises four subsections. Subsection III.A presents the crystal structure and electronic band structures of CaAgBi. Subsection II.B discusses the coexistence of type-I and type-II Weyl points  as well as topological surface states. Subsection III.C investigates Dirac and Weyl points using the low energy effective $k \cdot p$ models. Subsections III.D and E investigate and discuss the tunability of Dirac and dual-type Weyl points by alloy engineering in  $\mathrm{CaAgBi}_x\mathrm{Sb}_{1-x}$ and biaxial strain. Subsection III.F discusses an alternative candidate material CaAuBi under pressure. Concluding remarks are summarized in Section IV.

\section{\label{sec:level1}METHODOLOGY}

To investigate the crystal structure and electronic properties of CaAgBi, the first-principles calculations were conducted using density functional theory (DFT), as implemented in the Vienna Ab Initio Simulation Package (VASP) \cite{kresse1996efficient,kresse1999ultrasoft}, employing the projector-augmented wave (PAW) method \cite{blochl1994projector}. We used the Perdew–Burke–Ernzerhof (PBE) functional within the generalized gradient approximation (GGA) for the exchange-correlation functional \cite{perdew1996generalized}. A plane-wave basis set with a kinetic energy cutoff of 500 eV was used in all calculations.
The Brillouin zone was sampled with a Monkhorst-Pack $\mathit{k}$ point mesh \cite{monkhorst1976special} of size $9\times9\times7$ centered at the $\Gamma$ point. The structural optimization was performed until the forces on each atom were less than $0.01\,\mathrm{eV}/\mathrm{\AA}$. After full structural relaxation, we obtained the lattice constants as $a = b = 4.894\,\mathrm{\AA}$ and $c = 7.882\,\mathrm{\AA}$, which are in good agreement with the experimental values \cite{sun2007synthesis}($a = b = 4.811\,\mathrm{\AA}$ and $c = 7.827\,\mathrm{\AA}$). To explore the topological properties of CaAgBi, we constructed a tight-binding model and Hamiltonian based on DFT results using the maximally localized Wannier function (MLWF) \cite{mostofi2014updated} for Ca-d, Ag-s and Bi-p orbitals with the Wannier90 software \cite{mostofi2008wannier90}. Fermi arcs were calculated by using the recursive Green’s function method \cite{sancho1985highly} based on MLWF by WannierTools \cite{WU2018}.

\section{\label{sec:level1}Results and discussion}
\subsection{Crystal and electronic structures}
\begin{figure}[tp]
    \centering
    \includegraphics[width=1\linewidth]{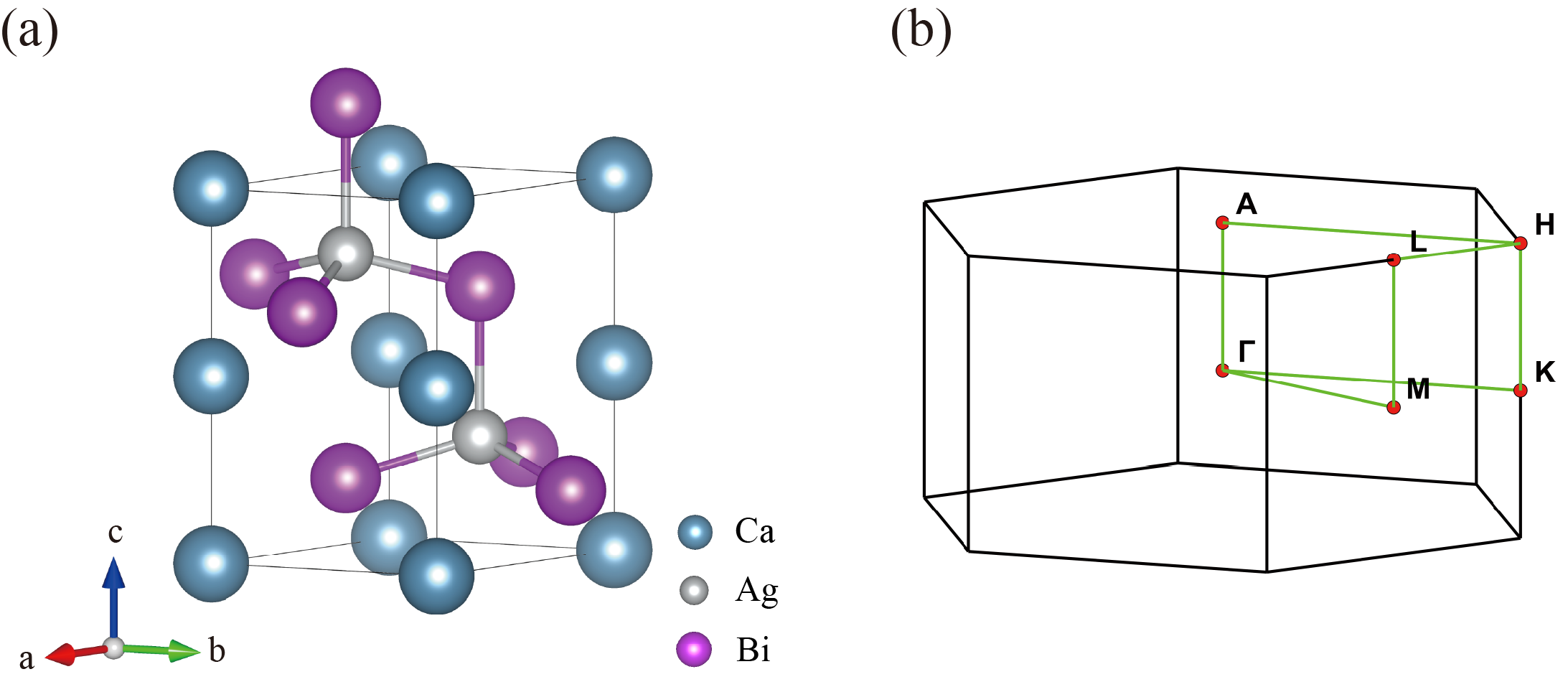}
    \caption{(a) The crystal structure of hexagonal CaAgBi. The blue, grey, and purple balls indicate Ca, Ag, and Bi atoms, respectively. (b) Brillouin zone and high symmetry points and paths.}
    \label{fig:1}
\end{figure}

As shown in Fig.~\ref{fig:1}, the crystal structure and Brillouin zone schematic of CaAgBi are presented. It crystallizes in a noncentrosymmetric Wurtzite-like structure with Ag and Bi forming the tetrahedral framework and Ca atoms occupying interstitial sites. The space group of CaAgBi is  $\mathit{P}6_3\mathit{mc}$ which corresponds to the $\mathit{C}_{6v}$ point group. It contains mirror symmetry $\mathit{M}_{yz}$, threefold rotational symmetry $\mathit{C}_{3z}$, and twofold screw symmetry $\mathit{S}_{2z}$. The buckling layer formed by Ag and Bi atoms breaks inversion symmetry, which makes the existence of Weyl points possible in this structure. 

\begin{figure}[bp]
    \centering
    \includegraphics[width=1\linewidth]{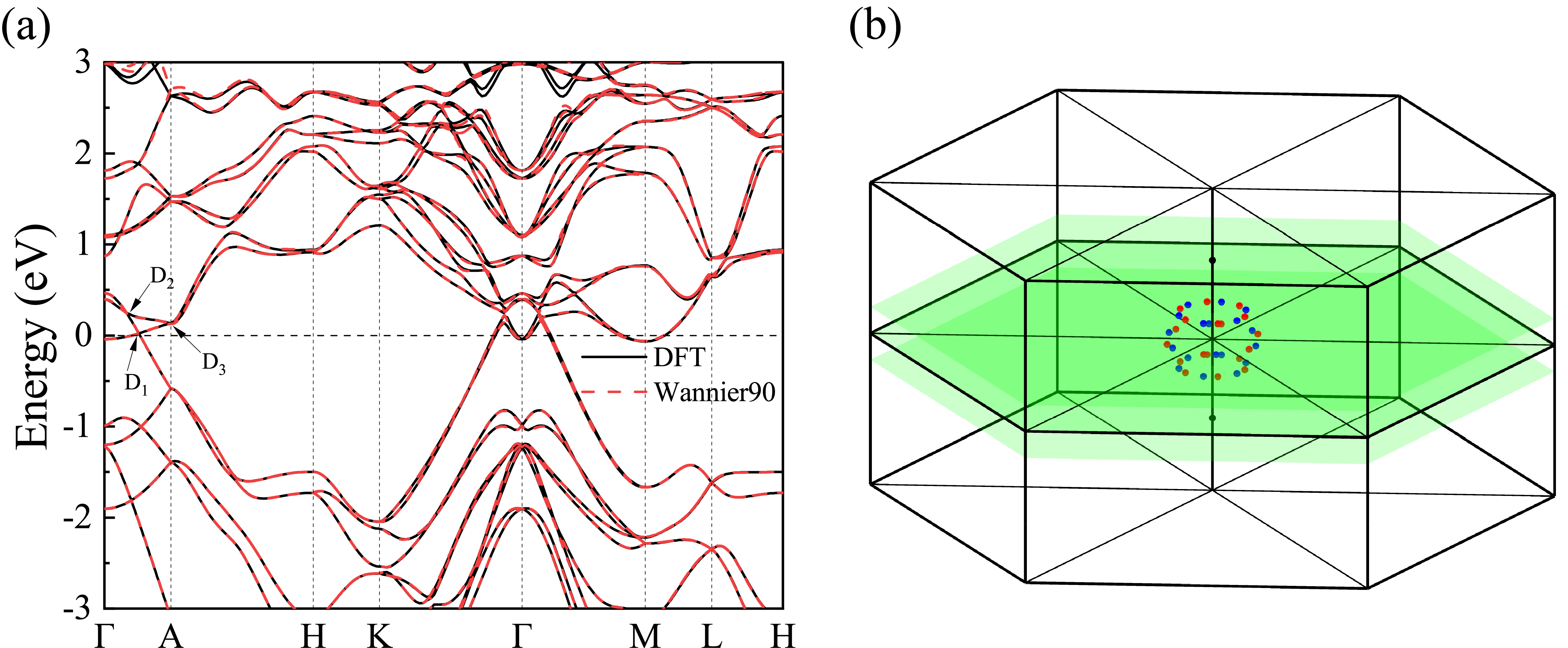}
    \caption{(a) Band structure by DFT calculations (black solid lines) and tight-binding model based on MLWF (red dashed lines) of CaAgBi.  $\mathrm{D_1}$, $\mathrm{D_2}$, and $\mathrm{D_3}$ indicate Dirac points. (b) Distribution of Dirac points (black dots) and Weyl points (blue and red dots) in the Brillouin zone.}
    \label{fig:2}
\end{figure}

The band structure in presence of spin-orbit coupling (SOC) along the high-symmetry lines of the Brillouin zone is shown in Fig.~\ref{fig:2}(a). There are three band crossing points along the $\Gamma–\mathrm{A}$ high-symmetry line. Although the system lacks inversion symmetry, the band crossings along this high symmetry line remain degenerate due to the combined protection of $\mathit{C}_{3z}$ (or $\mathit{S}_{2z}$) and $\mathit{M}_{yz}$ symmetries, preventing SOC-induced gap opening. This indicates that all three points are fourfold-degenerate Dirac points. Among the three points, $\mathrm{D_1}$ and $\mathrm{D_2}$ are accidental Dirac points arising from band inversion, similar to those found in $\mathrm{Na_3Bi}$ \cite{liu2014discovery} and SrHgPb \cite{gao2018dirac}. In contrast, $\mathrm{D_3}$ located at a high symmetry point in the Brillouin zone is an essential Dirac point, consistent with previous results \cite{chen2017ternary}. To explore the position of Weyl points and the topological surface state of CaAgBi, we constructed the tight-binding model based on MLWF and the corresponding band structure is shown as the red dashed line in Fig.~\ref{fig:2} (a). It turns out that our tight-binding model exhibits good agreement with the DFT-based band structure within the energy range from $-3\,\mathrm{eV}$ to $3\,\mathrm{eV}$.

Since CaAgBi crystallizes in the same space group as Dirac-Weyl semimetal SrHgPb, similar electronic structures and topological properties are anticipated. To verify this, we systematically searched for band crossings near the Fermi level across the Brillouin zone using a Wannier-function-based tight-binding model. It turns out that there are two Dirac points and 36 Weyl points, and their distribution in the Brillouin zone is shown in Fig.~\ref{fig:2} (b). The black points located on the $k_z$ axis represent Dirac points, corresponding to $\mathrm{D_1}$ in the band structure. These points exist along the $\Gamma–\mathrm{A}$ line at $k_z = \pm 0.209 \, \frac{2\pi}{c}$
, with a corresponding energy of $E-E_f=0.0197 \,\mathrm{eV}$. Additionally, we identified six pairs of Weyl points in each of three distinct planes: the $k_z=0$ and $k_z=\pm 0.0698\,\frac{2\pi}{c}$ planes, yielding a total of 18 pairs. Their energies are $E-E_f=0.2125 \,\mathrm{eV}$ and $E-E_f=0.2118\, \mathrm{eV}$. This configuration—comprising 18 Weyl pairs—contrasts markedly with SrHgPb \cite{gao2018dirac} (six pairs exclusively $k_z=0$) and MoTe/WTe \cite{meng2019dirac} (twelve pairs in $k_z\neq 0$  planes). Notably, the momentum-space distribution bears a closer resemblance to TaAs \cite{sun2015topological}.

Figs.~\ref{fig:3}(a) and \ref{fig:3}(b) display the band structures along selected paths in the $k_z=0$ and $k_z= 0.0698\,\frac{2\pi}{c}$ planes, respectively. The path endpoints correspond to identified Weyl points at coordinates: k=(0.0154, 0.119, 0) (Weyl point1) and k=(0.0644, 0.0579, 0.0698) (Weyl point2). Band crossings at $E-E_f =0.2125\, \mathrm{eV}$ in Fig.~\ref{fig:3}(a) and $E-E_f=0.2118\, \mathrm{eV}$ in Fig.~\ref{fig:3}(b) align precisely with the tight-binding predictions—validating the Weyl point positions in momentum space. Crucially, these band crossing points exhibit twofold degeneracy, distinct from Dirac points on high-symmetry lines, due to the absence of rotational symmetry protection away from the $k_z$ axis.

\subsection{Topological dual-type Weyl points and surface states}
\begin{figure}[tp]
    \centering
    \includegraphics[width=1\linewidth]{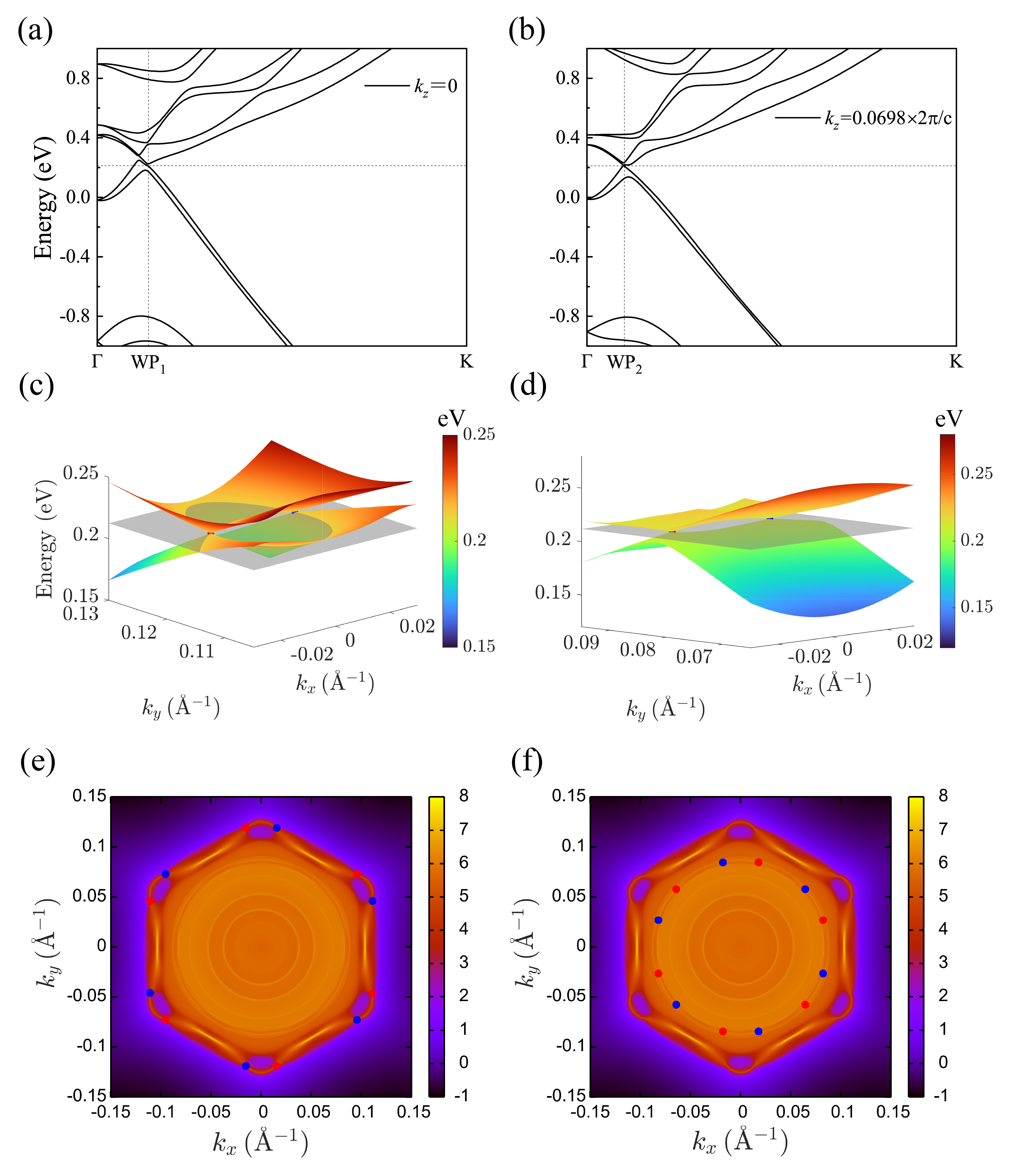}
    \caption{Band structures along $\Gamma $-$ \mathrm{WP} $-$ \mathrm{K}$ in the (a) $k_z=0$ plane and (b) $k_z=0.0698\,\frac{2\pi}{c}$ plane. (c) 2D band structures near the Weyl points in $k_z=0$ plane and (d) $k_z=0.0698 \,\frac{2\pi}{c}$ plane. The density of (001) surface states with (e)  $E-E_f=0.2125\,\mathrm{eV}$ and  (f)  $E-E_f=0.2118\,\mathrm{eV}$. The red and blue points represents the surface projected Weyl points with positive and negative chirality, respectively. }
    \label{fig:3}
\end{figure}

Weyl points can be classified into two fundamental types: type-I Weyl points feature a point-like Fermi surface with Lorentz-symmetric, isotropic conical dispersion and uniform Fermi velocity; type-II Weyl points exhibit a strongly tilted cone, yielding a line-like Fermi surface and Lorentz-symmetry violation due to anisotropic dispersion \cite{soluyanov2015type}. In CaAgBi, the Weyl points in the $k_z=0$ plane (Fig.~\ref{fig:3}(c)) exhibit a line-like Fermi surface characteristic of type-II Weyl points, while those at $ k_z = 0.0698\,\frac{2\pi}{c}$ plane (Fig.~\ref{fig:3}(d)) show the type-I point-like Fermi surface. Due to the time-reversal symmetry, the Weyl points located on the $k_z= -0.0698\,\frac{2\pi}{c}$ plane are also type-II. This constitutes the first observation of intrinsic type-I/II Weyl fermion coexistence in a Dirac-Weyl semimetal—previously reported only in externally tuned systems \cite{zhang2018coexistence,xu2020comprehensive}. Here, we do not find any explicit factor that determines the type of Weyl points, which appear to be accidental.

Surface electronic excitations in topological semimetals exhibit unique physical phenomena. In particular, Weyl semimetals host characteristic surface Fermi arcs on specific crystal surfaces. To determine the chirality of each Weyl point, we calculated the integral of Berry curvature over a spherical region with a radius of $0.005\,\mathrm{\AA}$ around each Weyl point. The results are presented in Fig.~\ref{fig:2}(b), with red indicating positive chirality and blue negative chirality. Based on the tight-binding model of CaAgBi, we applied the surface Green's function method to obtain the surface energy spectrum. The Fermi arc was identified on the projected (001) surface spectrum at $E-E_f=0.2125\,\mathrm{eV}$ as shown in Fig.~\ref{fig:3}(e). Furthermore, Weyl points with opposite chirality are connected pairwise by these Fermi arcs. In Fig.~\ref{fig:3}(f), the Fermi arcs projected from the $k_z= 0.0698\,\frac{2\pi}{c}$ planes are less pronounced compared to those from the $k_z=0$ plane. It can be observed that the surface states in Figs.~\ref{fig:3}(e) and ~\ref{fig:3}(f) are broadly similar, which can be attributed to the nearly identical energies of the Weyl points in the two planes. Notably, the separation between Weyl points in the $k_z=0$ plane is about $0.09 \,\mathrm{\AA}^{-1}$, larger than that in SrHgPb ($0.07 \,\mathrm{\AA}^{-1}$) \cite{gao2018dirac} and $\mathrm{MoTe}_2$ ($0.042 \,\mathrm{\AA}^{-1}$) \cite{sun2015prediction}. This wider separation indicates that these Weyl points may be more readily detectable by ARPES.

\subsection{Low energy effective Hamiltonian}
\begin{figure}[tp]
    \centering
    \includegraphics[width=1\linewidth]{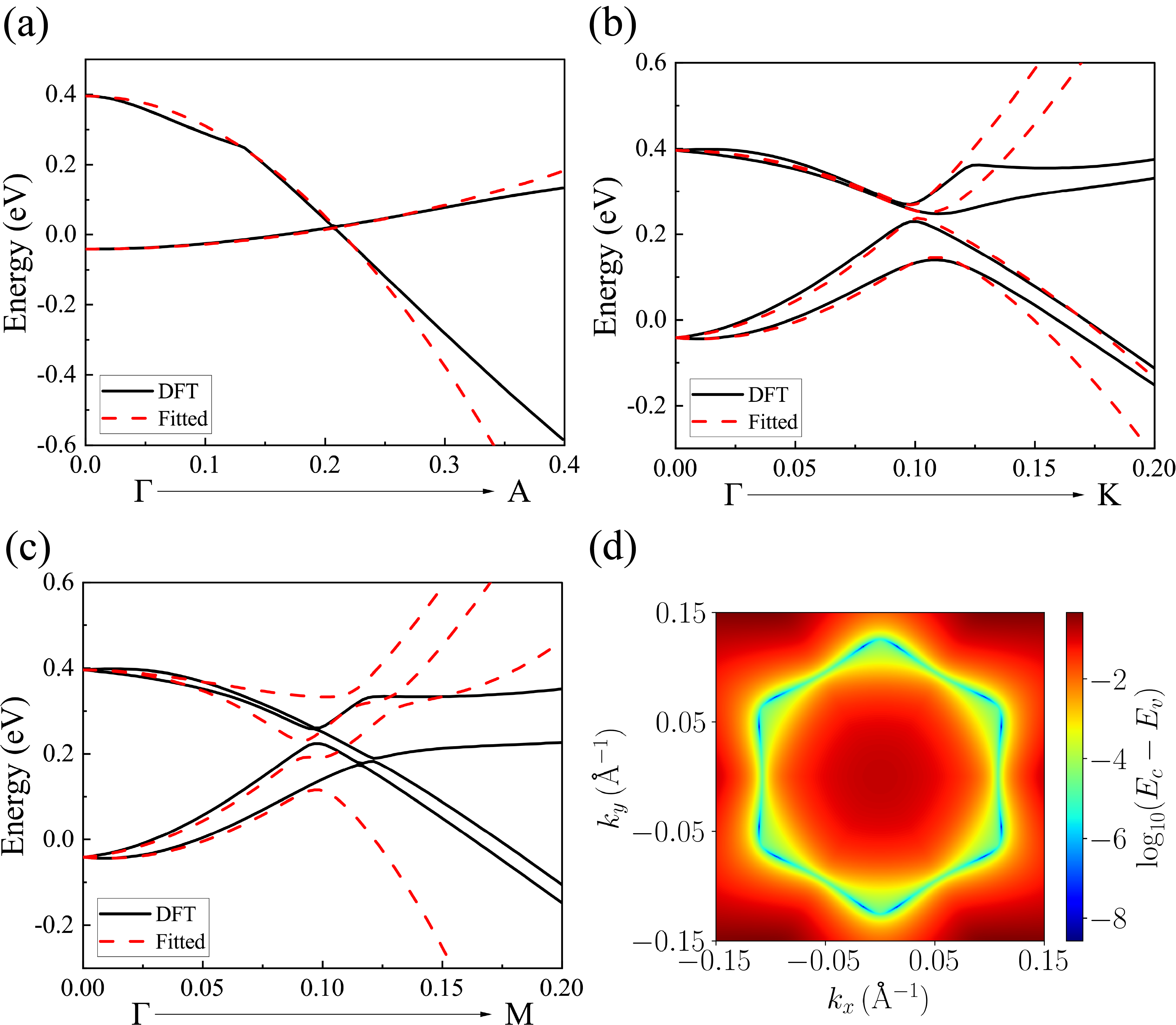}
 \caption{Comparisons for the band structures between DFT (black lines) and $k \cdot p $ model (red dash lines) along the high symmetry lines (a) $\Gamma-\mathrm{A}$ , (b)  $\Gamma-\mathrm{K}$ , and (c) $\Gamma-\mathrm{M}$, respectively. (d) Contour plot of energy difference between the conduction band and valence band, represented as $\log_{10}(E_c - E_v)$.}
    \label{fig:4}
\end{figure}

We can utilize a low-energy effective Hamiltonian to capture the Dirac and Weyl points in the band structure. To describe the low-energy electronic structure near the $\Gamma$ point, we introduce the following $4\times4$ effective Hamiltonian matrix model \cite{gao2018dirac}:
\begin{widetext}
\begin{equation}
H_{\mathbf{k}} = M_{\mathbf{k}} + 
\begin{pmatrix}
m_{\mathbf{k}} & ig_1 k_{-} + g_2 k_z k_{-} & ig_3 k_z k_{-}^{2} + g'_2 k_{-}^{2} & i\lambda_{\mathbf{k}} \\
-ig_1 k_{+} + g_2 k_z k_{+} & -m_{\mathbf{k}} & -ig'_1 k_{-} & -ig_3 k_z k_{-}^{2} + g'_2 k_{-}^{2} \\
-ig_3 k_z k_{+}^{2} + g'_2 k_{+}^{2} & ig'_1 k_{+} & -m_{\mathbf{k}} & ig_1 k_{-} - g_2 k_z k_{-} \\
-i\lambda_{\mathbf{k}}^{*} & ig_3 k_z k_{+}^{2} + g'_2 k_{+}^{2} & -ig_1 k_{+} - g_2 k_z k_{+} & m_{\mathbf{k}}
\end{pmatrix}
\end{equation}
\end{widetext}
where
$M_{\mathbf{k}} = M_0 + M_1(k_x^2 + k_y^2) + M_2 k_z^2$, $m_{\mathbf{k}} = m_0 -m_1(k_x^2 + k_y^2) - m_2 k_z^2$, $\lambda_{\mathbf{k}} = g'_3 k_+^3 + g''_3 k_-^3$, $k_{\pm} = k_x \pm i k_y$. The Hamiltonian matrix here takes into account the crystal symmetries $C_{6z}$, $M_{xz}$, and $T$. The basis states used are $\left| V_B, \tfrac{3}{2} \right\rangle$, $\left| C_B, \tfrac{1}{2} \right\rangle$, $\left| C_B, -\tfrac{1}{2} \right\rangle$, $\left| V_B, -\tfrac{3}{2} \right\rangle$.
We first fitted the band structure along the $\Gamma-\mathrm{A}$ direction, which determines the values of the parameters $M_0$ and $m_0$. It is found that the energy at the $\Gamma$ point is primarily controlled by these two parameters, reflecting the strength of intrinsic SOC. Further fitting reveals that the position of the Dirac points is primarily determined by the collective values of the parameters $M_0$, $m_0$, $M_2$, $m_2$. By fitting the band structures along the $\Gamma–\mathrm{A}$, $\Gamma–\mathrm{K}$, and $\Gamma–\mathrm{M}$ paths, we obtained the results shown in Figs.~\ref{fig:4}(a–c). It can be observed that the DFT results are in good agreement with the $k\cdot p$ model results. The remaining differences can be attributed to the approximation in the model, which includes terms only up to third order. In Fig.~\ref{fig:4}(d), the blue regions correspond well to the positions of the Weyl points. Since the $k\cdot p$ model is valid for low-energy regions and the vicinity of the $\Gamma$ point in momentum space, we did not perform fitting for the band structure at $k_z\neq0$ planes in this case. The fitted parameters are listed in Table~\ref{tab:fitted parameters}.

\begin{table}[h]
\caption{\label{tab:fitted parameters}%
The fitted parameters of the $k\cdot p$ Hamiltonian.
}
\begin{ruledtabular}
\begin{tabular}{ccccc}
\textrm{$M_0(\,\mathrm{eV})$}&
\textrm{$M_2(\,\mathrm{eV\AA^2})$}&
\textrm{$m_0(\,\mathrm{eV})$}&
\textrm{$m_2(\,\mathrm{eV\AA^2})$}&
\textrm{$M_1(\,\mathrm{eV\AA^2})$}\\
0.18 & -3.63 & 0.22 & 5.05 & 4.67\\

\textrm{$m_1(\,\mathrm{eV\AA^2})$}&
\textrm{$g_1(\,\mathrm{eV\AA})$}&
\textrm{$g'_1(\,\mathrm{eV\AA})$}&
\textrm{$g_2(\,\mathrm{eV\AA^2})$}&
\textrm{$g'_2(\,\mathrm{eV\AA^2})$}\\
18.93 & -0.42 & 0.096 & 23.67 & 3.46\\

\textrm{$g_3(\,\mathrm{eV\AA^3})$}&
\textrm{$g'_3(\,\mathrm{eV\AA^3})$}&
\textrm{$g''_3(\,\mathrm{eV\AA^3})$}\\
-3.25 & -38.45 & 49.16\\
\end{tabular}
\end{ruledtabular}
\end{table}

\subsection{Tunable Dirac and Weyl points by alloy engineering}
Modulating Weyl point positions in topological materials allows control over surface Fermi arcs, topological phase transitions, and electronic transport properties. Previous strategies include atomic position engineering in SrHgPb \cite{gao2018dirac} and uniaxial strain in MoTe(WTe) \cite{meng2019dirac}. In this work, we apply alloy doping to the isostructural CaAgBi–CaAgSb system, exploiting their contrasting topological (semimetal vs. trivial) properties as a tuning pathway, and find it provides great tunability to the system.

\begin{figure}[tp]
    \centering
    \includegraphics[width=1\linewidth]{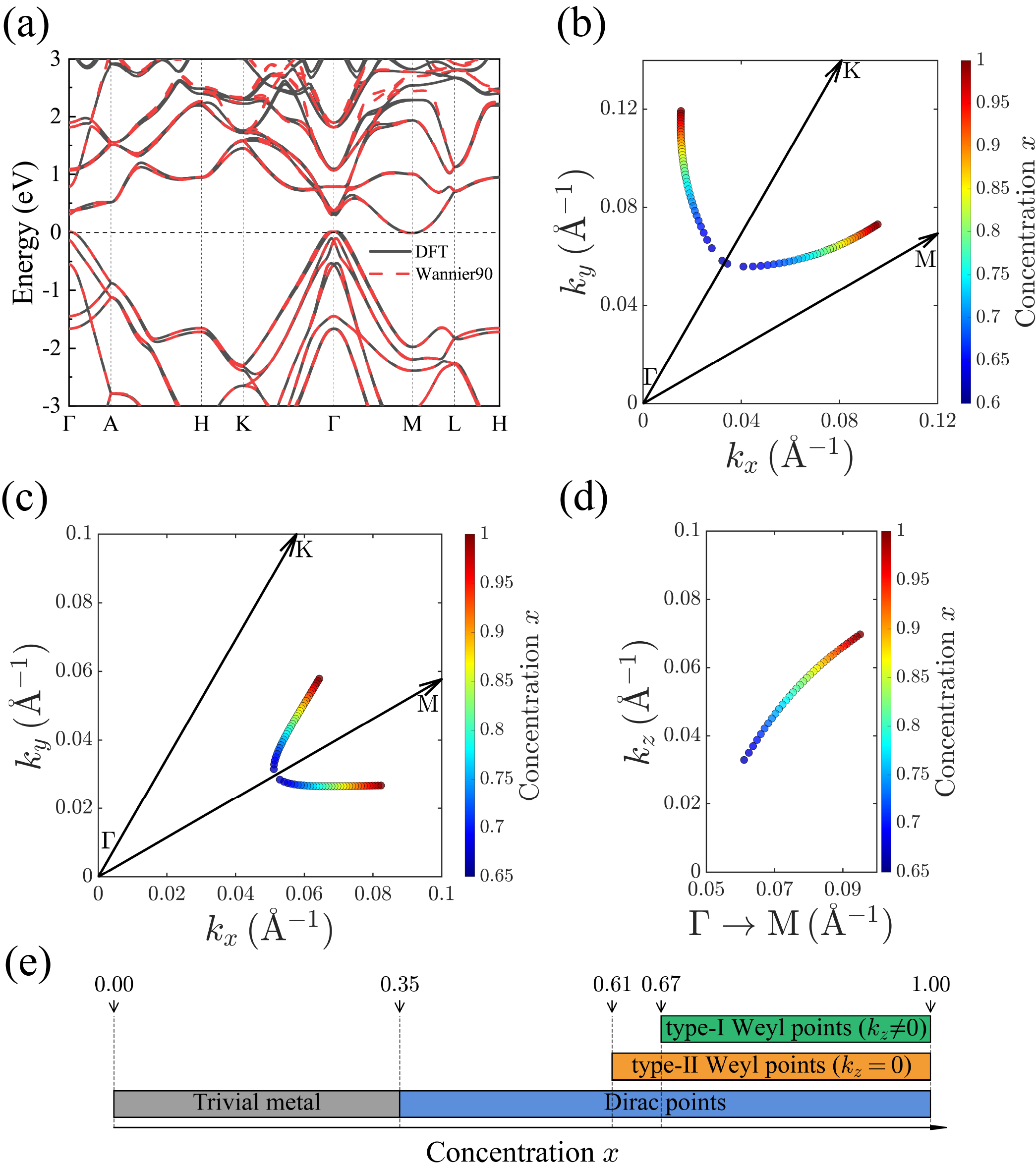}
    \caption{(a) Band structures of CaAgSb by DFT calculations (black solid lines) and tight-binding model based on MLWF (red dashed lines). (b) Evolution of Weyl point positions in the $k_z=0$ plane in  \(\mathrm{CaAg}\mathrm{Bi}_{x}\mathrm{Sb}_{1-x}\) alloy with $x$ from 0.6 to 1 . (c) and (d) Projected Weyl point positions in the $k_x$–$k_y$ plane and the $\Gamma$–$\mathrm{M}$–$k_z$ plane (for $k_z \ne 0$), respectively, in CaAgBi$_x$Sb$_{1-x}$ alloy with $x$ from 0.65 to 1. (e) Schematic diagram of the phase transition and types of Weyl points with concentration $x$.}
    \label{fig:5}
\end{figure}

Fig.~\ref{fig:5}(a) shows the first-principles band structure of CaAgSb with $P6_3mc$ symmetry. It turns out that CaAgSb is a topologically trivial metal due to the conduction band touching the Fermi level at the M point. We employed the virtual crystal approximation (VCA)~\cite{porod1983modification} to model the alloys, constructing MLWFs from Ca-d, Ag-s, and Sb-p orbitals, and obtained the Hamiltonian of $\mathrm{CaAgBi}_x\mathrm{Sb}_{1-x}$ by linear interpolation between those of CaAgBi and CaAgSb~\cite{gao2021noncentrosymmetric}.

Our calculations revealed the coexistence of Weyl and Dirac points within a Bi concentration range of $x$ from 0.61 to 1. The Dirac points are located along the $k_z$ axis. As the concentration $x$ decreases from 1 to 0.35, the $k_z$ values of Dirac points shift from $0.209\,\frac{2\pi}{c}$ to 0, leading to a transition into a trivial metal. It is noteworthy that the type-II Weyl points in the $k_z=0$ plane and type-I Weyl points in $k_z \neq 0$ planes vanish at different concentrations, as summarized in Fig.~\ref{fig:5}(e). Fig.~\ref{fig:5}(b) illustrates that the type-II Weyl points remain stable for $0.61 \leq x \leq 1$ but annihilate at $x=0.6$ on the $\Gamma$–K path. As shown in Fig.~\ref{fig:5}(c), type-I Weyl points in the $k_z \ne 0$ planes annihilate at $x=0.67$ along the $\overline{\Gamma}$–$\overline{\mathrm{M}}$ line ($k_z = \pm 0.0414\,\frac{2\pi}{c}$), leaving only type-II Weyl points for $0.61 \leq x \leq 0.67$. This range thus realizes a Dirac–Weyl semimetal phase similar to SrHgPb. Fig.~\ref{fig:5}(e) provides a schematic summary of the topological phase transition in $\mathrm{CaAgBi}_x\mathrm{Sb}_{1-x}$ alloy with $x$. Furthermore, Fig.~\ref{fig:5}(d) shows a decreasing trend in the $k_z$ of Weyl points with reduced Bi concentration. As evident from all three projections, Weyl points progressively approach the $\Gamma$ point as the Bi concentration decreases. This behavior can be attributed to the role of SOC: heavy Bi (Z=83) atoms introduce strong SOC into the system, while doping with lighter Sb (Z=51) atoms weakens this SOC effect, thus suppressing band inversion along the $\Gamma–\mathrm{M}$ and $\Gamma–\mathrm{K}$ lines. This weakening drives the Weyl points closer to the $\Gamma$ point until they eventually annihilate.

\subsection{Tunable Weyl points under biaxial strain}

\begin{figure}[tp]
    \centering
    \includegraphics[width=0.9\linewidth]{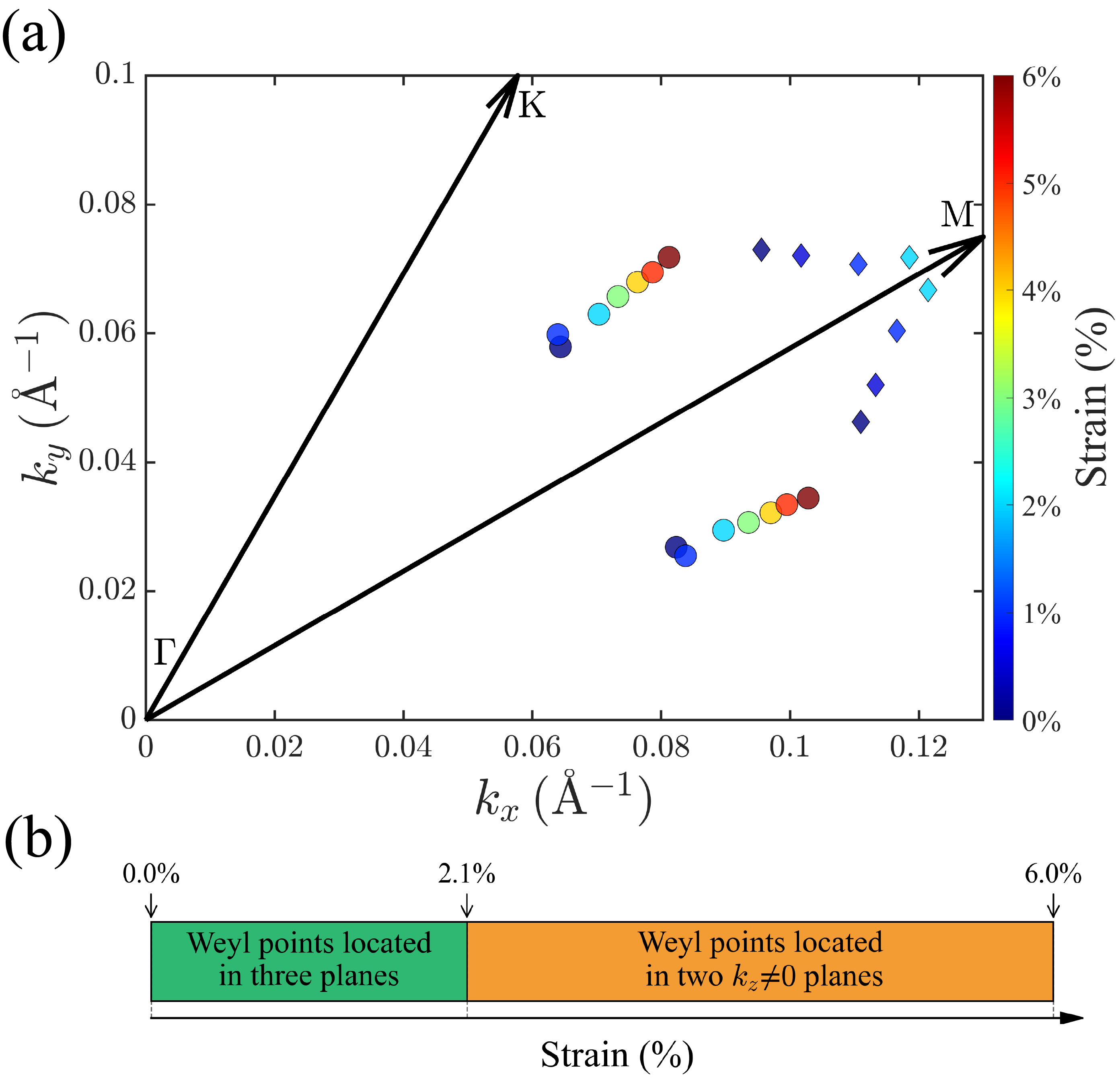}
    \caption{(a) Positions of projected Weyl points under biaxial strain. Dots represent Weyl points in the $k_z \ne 0$ planes, while diamonds represent  Weyl points in the $k_z = 0$ plane. (b) Schematic diagram of the distribution of Weyl points under biaxial strain.}
    \label{fig:6}
\end{figure}

We also investigated the stability of the Weyl and Dirac points in CaAgBi under biaxial tensile strain ranging from $ 0\% $ to $6\%$. Dirac points located at the $k_z$ axis remain stable, gradually shifting from 0.209 to $0.237\,\frac{2\pi}{c}$. As shown in Fig.~\ref{fig:6}(a), when the tensile strain reaches $2.1\%$, Weyl points in the $k_z  = 0$ plane (marked with diamond symbols) annihilate along the high-symmetry $\Gamma–\mathrm{M}$ path, indicating that these Weyl points are unstable under tensile strain. Meanwhile, these type-II Weyl points transform into type-I when the strain exceeds $1.96\%$, before eventually annihilating. In contrast, Weyl points located in the $k_z \neq 0$ planes (marked with dots) do not exhibit any tendency to annihilate. Instead, they exhibit an outward shift from the $\Gamma$ point as the in-plane lattice constants (ab-plane) increase with strain. The distribution of Weyl points under biaxial strain is summarized in Fig.~\ref{fig:6}(b). The results suggest that the Dirac-Weyl phase remains stable under tensile strain up to $6\%$, exhibiting greater robustness than SrHgPb (3.5\%). In addition to applying tensile strain, we examined the stability of the system under hydrostatic pressures up to 10 GPa, and found that the Dirac-Weyl phase remains robust against external perturbations.

\subsection{Other material candidate: CaAuBi}

\begin{figure}[tp]
    \centering
    \includegraphics[width=1\linewidth]{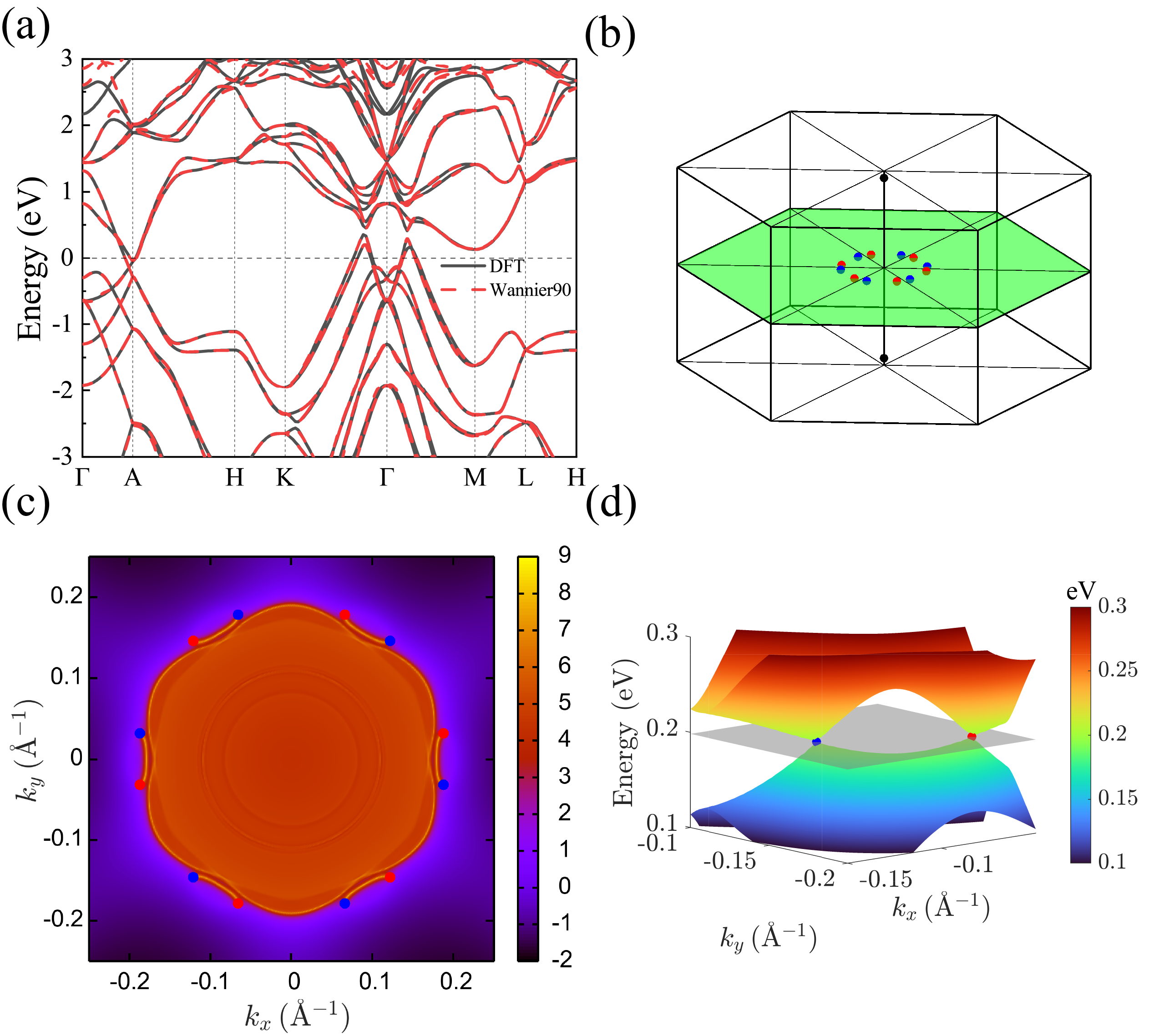}
    \caption{(a) Band structures by DFT calculations (black solid lines) and tight-binding model based on MLWF (red dashed lines) of CaAuBi. (b) Distribution of Dirac points (black dots) and Weyl points (blue and red dots) in the Brillouin zone for CaAuBi. (c) The density of (001) surface states at  $E - E_f  = 0.1984\,\mathrm{eV}$. (d) 2D band structures near the Weyl points.}
    \label{fig:7}
\end{figure} 

We also find that CaAgBi family material: CaAuBi, in its high-pressure phase ($P6_3mc$) \cite{xie2014pressure}, exhibits the characteristics of a Dirac–Weyl semimetal.

Using the method described above, we get the results below. The overall band structure in Fig.~\ref{fig:7}(a) resembles that of known topological semimetals such as CaAgBi and SrHgPb, featuring a clear band inversion along the high-symmetry path $\mathrm{K}$--$\Gamma$--$\mathrm{M}$ and $\Gamma$--$\mathrm{A}$. Although the Weyl points in the $k_z=0$ plane of SrHgPb and CaAgBi are both type-II, the point-like Fermi surface structure around the Weyl nodes in Fig.~\ref{fig:7}(d) indicates that these Weyl points are of type-I.

\section{\label{sec:level1}CONCLUSION}
In this work, we systematically investigate the topological properties of CaAgBi with the $\mathit{P}6_3\mathit{mc}$ space group using first-principles calculations, Wannier-based tight-binding models, and a low-energy effective Hamiltonian. Distinct from conventional Dirac and Weyl semimetals, CaAgBi hosts both type-I and type-II Dirac (Weyl) points. Such coexistence has not been reported in other topological materials. Notably, the $k_z=0$ plane exhibits clear Fermi arcs with lengths of about $0.09 ,\mathrm{\AA}^{-1}$, suggesting that they can be readily observed by ARPES. To gain deeper insight into this Dirac–Weyl coexisting phase, we developed a $k\cdot p$ model around the $\Gamma$ point, which captures the key features of this unique topological phase. In addition, we tuned the topological properties through alloy doping and external strain, demonstrating both the continuous tunability of the system and the robustness of the Dirac–Weyl phase in CaAgBi. In summary, our study establishes CaAgBi as a robust and tunable platform for exploring interactions between distinct types of Dirac and Weyl fermions, and provides possible strategies for their manipulation.

\begin{acknowledgments}

This work was supported by the National Natural Science Foundation of China (Grants No. 12204299, No. 12074241, No. 11929401, No. 52130204, No. 12274278, and No. 12304340), the Shanghai Pujiang Program (Grant No. 23PJ1403200)
\end{acknowledgments}

\bibliography{refer}
\end{document}